\documentclass{article}
\usepackage[utf8]{inputenc}
\usepackage{amsmath}
\usepackage{graphicx}
\usepackage{float}
\usepackage{flexisym}
\usepackage[a4paper, total={6in, 9in}]{geometry}
\usepackage{multicol}
\usepackage{multirow}
\usepackage{multicol}
\usepackage[english]{babel}
\usepackage[backend=biber,style=nature,sorting=none]{biblatex}
\usepackage{authblk}
\usepackage[dvipsnames]{xcolor}

\title{\textbf{Overcoming Boltzmann's Tyranny in a Transistor via the Topological Quantum Field Effect}}
\author[1,2]{Muhammad Nadeem}
\author[3,4]{Iolanda Di Bernardo}
\author[1,2]{Xiaolin Wang}
\author[3,4]{Michael S. Fuhrer}
\author[5,6]{Dimitrie Culcer}
\affil[1]{Institute for Superconducting and Electronic Materials (ISEM), Australian Institute for Innovative Materials (AIIM), University of Wollongong, Wollongong, New South Wales 2525, Australia.}
\affil[2]{ARC Centre of Excellence in Future Low-Energy Electronics Technologies (FLEET), University of Wollongong, Wollongong, New South Wales 2525, Australia}
\affil[3]{School of Physics and Astronomy, Monash University, Clayton, Victoria 3800, Australia}
\affil[4] {ARC Centre of Excellence in Future Low-Energy Electronics Technologies (FLEET), Monash University, Clayton, Victoria 3800, Australia.}
\affil[5]{School of Physics, University of New South Wales, Sydney 2052, Australia}
\affil[6]{ARC Centre of Excellence in Future Low-Energy Electronics Technologies (FLEET), University of New South Wales, Sydney 2052, Australia}

\date{}

\begin{document}
\maketitle
The sub-threshold swing is the fundamental critical parameter determining the operation of a transistor in low-power applications such as switches. It determines the fraction of dissipation due to the gate capacitance used for turning the device on and off, and in a conventional transistor it is limited by \textit{Boltzmann's tyranny} to $k_BTln(10)/q$, or 60 mV per decade. Here, we demonstrate that the sub-threshold swing of a topological transistor, in which conduction is enabled by a topological phase transition via electric field switching, can be sizably reduced in a \textit{non-interacting} system by modulating the Rashba spin-orbit interaction via a top-gate electric field. We refer to this as the \textit{Topological Quantum Field Effect} and to the transistor as a \textit{Topological Quantum Field Effect transistor} (TQFET). By developing a general theoretical framework for quantum spin Hall materials with honeycomb lattices we explicitly show that the Rashba interaction can reduce the sub-threshold swing by more than 25\% compared to Boltzmann's limit in currently available materials, but without any fundamental lower bound, a discovery that can guide future materials design and steer the engineering of topological quantum devices.\par

\begin{flushleft}
\textbf{Introduction}\\
\end{flushleft}
A large fraction of power dissipation in the low-energy operation of a conventional semiconductor transistor occurs due to irreversible charging and discharging of the gate capacitor to turn conduction on and off. Its efficiency is characterized by the sub-threshold swing, such that a transistor with a small sub-threshold swing transitions rapidly between its on (high current) and off (low current) states. Yet in a MOSFET this swing is restricted by the fundamental limit $k_BT\ln(10)/q$, frequently termed \textit{Boltzmann's tyranny}, attained when the gate capacitance is infinitely large. This value can be understood by noting that a gate voltage $V$ raises a barrier $qV$ in the channel of a MOSFET, while the current is determined by the number of carriers which thermally activate over the barrier. For $qV \gg  k_BT$ the current $I \propto e^{-k_BT/qV}$, hence the sub-threshold swing $S = [d \log(I)/dV]^{-1} = k_BT\ln(10)/q$. \par

Strategies to lower the sub-threshold swing of next-generation transistors resort to either tunnelling or electron-electron interactions, whether in the gate capacitor \cite{Kobayashi18} or in the channel \cite{Newns00,Banerjee09}. The non-linear dielectric response of ferroelectric insulators can be harnessed to design gate insulators with a negative capacitance \cite{Iniguez19}, which increases the bending of the surface potential as a function of the top gate potential, yet this regime tends to be energetically unstable \cite{collaert2018}. Tunneling FETs \cite{Ionescu11} rely on charge tunnelling between spatially separated valence and conduction bands as a source of carrier injection within a pre-determined energy window, yet tunnelling transport restricts the current in the ON-state to relatively low values. 

In this work we address the generic problem of the energy gap $E_G$ to transport opened by a gate potential $U_G=qV$ applied to an electrode and demonstrate that Boltzmann’s tyranny can be overcome in a non-interacting electron system. We define a reduced sub-threshold swing as $S^* = [dE_G/d(U_G)]^{-1}$, such that the full swing $S = S^* k_B T\ln(10)/q$, and for a MOSFET in general $[dE_G/d(U_G)]^{-1} \ge 1$, so $S \ge k_BT\ln(10)/q$. For a gapped system described by a single-particle Hamiltonian in the presence of terms linear in the applied potential, with sub-unitary proportionality constants, it has hitherto been assumed that $S^* \ge 1$. Our work overturns this conventional wisdom, demonstrating that $S^*$ can be sub-unitary and is unbounded (though positive), implying that $S$ itself can be smaller than the value predicted by Boltzmann's tyranny in a general class of devices we refer to as topological transistors.

The past decade has shown that the topological properties of several classes of materials can lead to regimes in which transport is dissipationless. A new blueprint for transistor design has emerged in which conduction is turned on and off via a topological phase transition induced by a gate electric field \cite{WaryL, Ezaw13APL, Liu14-natmat, Liu15-nano, pan15sci, Qian14, zhange-natnano17,Molle-natmat17, Collins18}, that is, conventional carrier inversion is replaced by a topological phase transition. Whereas charge transport inside a topological transistor can be dissipationless, performance will be limited by the power dissipated in switching the transistor on and off. In the ideal case of fully dissipationless transport this will account for \textit{all} the dissipation in the transistor. In this paper we demonstrate that this process can be made energy efficient by resorting to the Rashba spin-orbit interaction (SOI) induced by a gate electric field \cite{Kane05a, Kane05b, Rashba09}. We refer to this mechanism as the topological quantum field effect, stressing that it has no counterpart in conventional MOSFETs, and to the transistor itself as a TQFET. This effect requires both the relativistic quantum mechanical phenomenon of spin-orbit coupling and proximity to a topological phase transition. Starting from a generic model describing the quantum spin Hall effect (QSHE) \cite{Kane05a, Kane05b}, we derive an expression for the subhreshold swing in the presence of Rashba SOI and apply it to materials with a honeycomb lattice structure \cite{HongkiMin06, LiuPRL11, LiuPRB11, XuPRL13, Hsu15, Reis17, Li18} that are being actively investigated. We find that the sub-threshold swing can be reduced by more than 25\% ($S^*<0.75$) compared to Boltzmann's limit in existing topological devices, and emphasize that it has no fundamental lower bound (though positive) and could be further improved by targeted materials design. \par

The reduction of the sub-threshold swing of a TQFET via Rashba interaction is closely associated with the microscopic bulk band topology stemming from the SOI. Instead of gate induced macroscopic current control, Rashba SOI effectively enhances the gate-induced topological phase transition by controlling the quantum dynamics at the microscopic scale, and hence the macroscopic edge state conductance through the bulk-boundary correspondence. Due to its dependence on the gate electric field, atomic SOI, geometric structure of QSH lattices and the Slater-Koster inter-orbital hopping parameters \cite{Slater-Koster}, the Rashba SOI provides tunable parameters for controlling the sub-threshold swing in a TQFET – rather than relying purely on the gate capacitance mechanism.

\begin{figure}[H]
	\centering
	\includegraphics[scale=0.85]{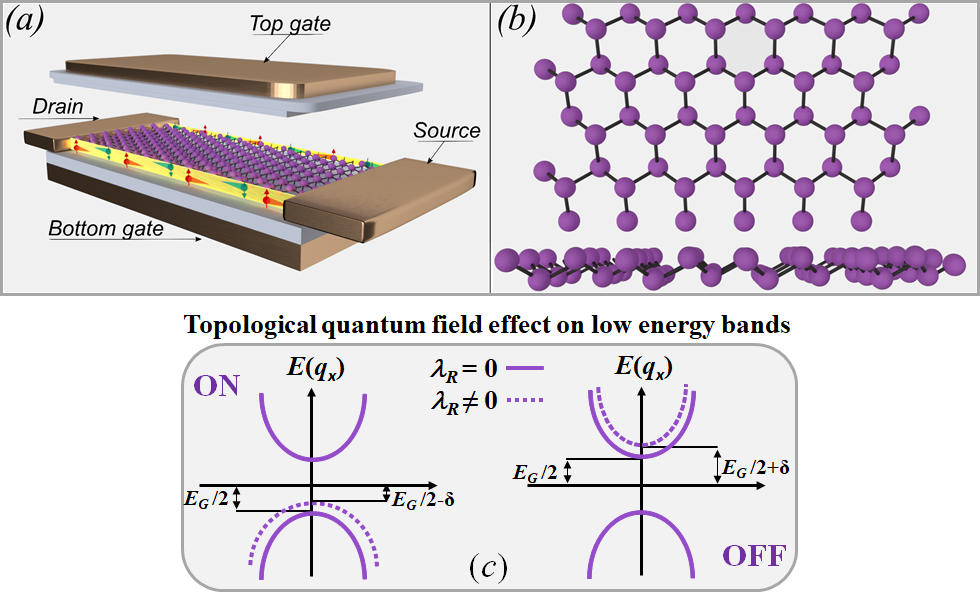}
	\caption{\textbf{Topological quantum field effect transistor}. In the absence of a gate electric field, a QSH insulator hosts dissipationless helical conducting channels with a minimum value of the quantized conductance $2e^2/h$ (ON state of TQFET (a)). When the gate electric field exceeds a threshold limit, the thin QSH insulator layer (staggered honeycomb lattice (b)) enters into the trivial regime, in which the minimum value of the conductance drops to zero (OFF state). Such electric field switching is accompanied by the topological quantum field effect which enhances the topological phase transition driven by a gate electric field and reduces the sub-threshold swing (c). Here, $\delta$ represents the shift in nontrivial/trivial band gap $E_G$ due to topological quantum field effect.}
\label{TQFET}
\end{figure}

\begin{flushleft}
\textbf{Model of a topological quantum field effect transistor}\\
\end{flushleft}
A transistor based on topologically nontrivial condensed matter systems hosting the QSH effect can be engineered via a topological phase transition induced by a gate electric field as shown in Figure \ref{TQFET}. Near energy-zero, the edge state conductance of QSH lattices in a nanoribbon geometry is (i) quantized, (ii) topologically protected, and (iii) associated with microscopic quantum phenomena in the bulk. The minimum value of the quantized conductance drops from $2e^2/h$ to zero as the system transits from the QSH phase (ON) into the trivial regime (OFF) via electric field switching. \par

The QSH has been proposed in graphene \cite{Kane05a,Kane05b} and other group-IV and V honeycomb lattice structures \cite{HongkiMin06,LiuPRL11,LiuPRB11,XuPRL13,Hsu15,Reis17,Li18}, monolayer transition metal dichalcogenides in the 1T\textprime configuration \cite{Qian14}, thin films of 3D topological insulators Bi\textsubscript{2}Se\textsubscript{3} \cite{Shan10,SCZhnag} as well as the Dirac semimetal Na\textsubscript{3}Bi \cite{Collins18}. The experimental demonstration of a very large nontrivial band gap of 360 meV for Na\textsubscript{3}Bi \cite{Collins18} and 800 meV for Bi/SiC \cite{Reis17}, significantly exceeding the thermal energy at room temperature (25 meV), indicates that QSH materials may be robust platforms for nano-electronic devices at room temperature. We note, in passing, that a small sub-threshold swing requires materials whose gap increases at a fast rate as a function of the top gate voltage, ruling out bilayer graphene, whose gap is found to increase more slowly than initial theoretical predictions \cite{McCann13}.

Consider a tight-binding model Hamiltonian reproducing the effective low energy Dirac theory specialized to QSH materials with honeycomb lattice structures \cite{Kane05a, Kane05b}
\begin{equation}
\begin{split}
    H & = t\sum_{\langle ij\rangle\alpha}c_{i\alpha}^\dagger c_{j\alpha} + i\lambda_{so}\sum_{\langle\langle ij \rangle\rangle\alpha\beta} v_{ij} c_{i\alpha}^\dagger s_{\alpha\beta}^z c_{j\beta}\\
    & +\frac{\lambda_{v}}{2}\sum_{\langle ij\rangle\alpha} c_{i\alpha}^\dagger v_{i} c_{j\alpha} + i\lambda_{R}(E_{z})\sum_{\langle ij\rangle\alpha\beta} c_{i\alpha}^\dagger (\textbf{s}_{\alpha\beta} \times \hat{\textbf{d}}_{ij})_{z} c_{j\beta}
\end{split}
\end{equation}
Here $c_{i\alpha}^\dagger (c_{i\alpha})$ is the creation (annihilation) electron operator with spin polarization $\alpha=\uparrow,\downarrow$ on site \textit{i}, the Pauli matrix $s^z$ describes the electron intrinsic spin while $s_{\alpha\beta}^z$ are the corresponding matrix elements describing the spin polarization $\alpha$ and $\beta$ on sites \textit{i} and \textit{j}, $v_{i}=+1(-1)$ for sublattice A (B), and $v_{ij}=\textbf{d}_{ik}\times\textbf{d}_{kj}=\pm1$ connects sites \textit{i} and \textit{j} on sublattice A (B) via the unique intermediate site \textit{k} on sublattice B (A). The nearest-neighbour bond vectors $\textbf{d}_{ik}$ and $\textbf{d}_{kj}$ connect the \textit{i} (\textit{k}) and \textit{k} (\textit{j}) sites on the A and B sublattices. The first term is the nearest neighbour hopping with amplitude \textit{t} while the second term is the intrinsic atomic Kane-Mele type SOI of strength $\lambda_{so}$. The third and fourth terms represent the staggered sublattice potential $\lambda_v=U_G$ and spin-mixing Rashba SOI associated with an externally applied gate electric field, $\lambda_{R}(0)=0$. Since the band gap closes at $q=0$ during the topological phase transition induced by the electric field, we ignore the $nnn$-intrinsic Rashba SOI, which does not affect the QSH band gap. In the long wavelength limit, with basis $\psi_{k}=\{a_{k,\uparrow},b_{k,\uparrow},a_{k,\downarrow},b_{k,\downarrow}\}$, the low energy effective four-band Bloch Hamiltonian H(\textbf{q}) in the vicinity of Dirac points K(K\textprime) reads
\begin{equation}
\begin{split}
    H(\textbf{q}) & = v_{F}s_{0}\otimes(\eta_{\tau}q_{x}\sigma_{x}+q_{y}\sigma_{y}) +\eta_{\tau}\Delta_{so}s_{z}\otimes\sigma_{z}\\
    & +(\Delta_{v}/2) s_{0}\otimes\sigma_{z} +\Delta_{R}(\eta_{\tau}s_{y}\otimes\sigma_{x}-s_{x}\otimes\sigma_{y})
\end{split}
\end{equation}
where $\eta_{\tau}=+(-)$ is the valley index representing K(K\textprime), $v_{F}=3at/2$ is the Fermi velocity where $a$ is the lattice constant, \textbf{\textit{s}} and $\boldsymbol{\sigma}$ are the spin and pseudospin Pauli matrices respectively, $\Delta_{v}=\lambda_{v}$ and $\Delta_{so}=3\sqrt{3}\lambda_{so}$ and $\Delta_{R}=3\lambda_{R}/2$ are SOI parameters. As shown in figure \ref{TQFET}, $\Delta_{v}=+U_G$ (or $ \Delta_{v}=2U_G$) for sublattice A and $\Delta_{v}=-U_G$ (or $\Delta_{v}=0$) for sublattice B in the dual-gate (or top-gate only) version. The dual gate version is equivalent to a simpler single top gate formulation where top gate adds positive potential terms only, so the A and B sublattices would be at $U_G$ and 0 respectively, which corresponds to applying $+U_G/2$ and $-U_G/2$ with a rigid shift of $U_G= ed_zE_z$. Here $-e$ is the electron charge, $E_{z}$ is the gate electric field, and $d_{z}$ is the distance of the $i^{th}$ site from the zero electric potential site.\par

\begin{figure}[H]
    \centering
    \includegraphics[scale=0.75]{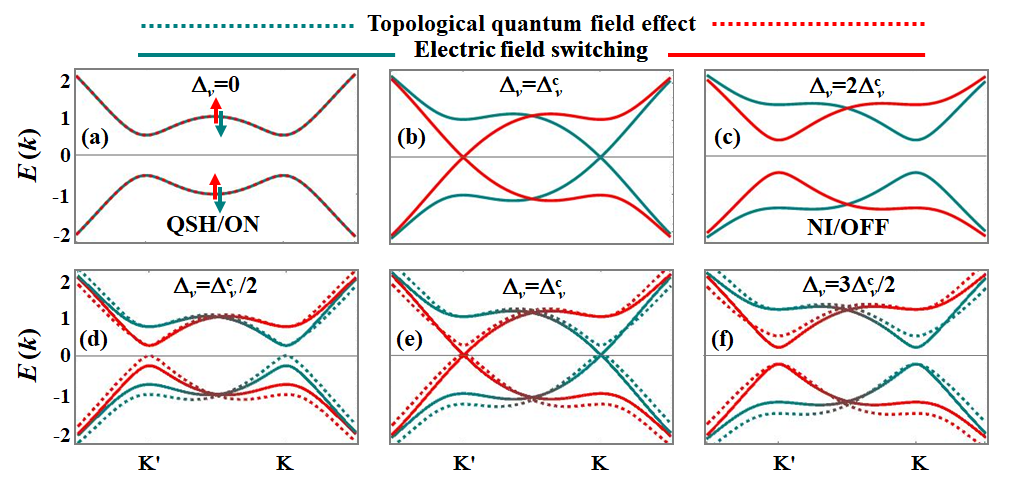}
    \caption{\textbf{Electric field switching $(\Delta_R=0)$ and Topological quantum field effect $(\Delta_R\ne0)$.} \textbf{(a-c)} In the absence of Rashba SOI, a uniform out-of-plane electric field drives QSH (a) to normal insulating (NI) phase (c) while passing through a critical point ($\Delta_{v}=\Delta_{v}^c$) where valleys K(K\textprime) are perfectly spin-polarized hosting spin down (up) gapless states (b). \textbf{(d-f)} The topological quantum field effect reduces the nontrivial band gap (d), opens the trivial band gap at the threshold gate electric field (e), and enhances the trivial band gap (f). The topological quantum field effect speeds up the topological phase transition by opening a trivial band gap when  $\Delta_{v}=\Delta_{v}^c$ (e) which would otherwise be the critical point in the absence of topological quantum field effect (b). Here we use $t=1$ eV, $\Delta_{so}=0.519$ eV, and $\Delta_{R}=0.225$ eV (d), $\Delta_{R}=0.30$ eV (e), $\Delta_{R}=0.375$ eV (f). The solid(dashed) lines represent band dispersion in the absence(presence) of Rashba SOI.}
    \label{TQFE-TBM}
\end{figure}

\begin{figure}[H]
    \centering
    \includegraphics[scale=0.8]{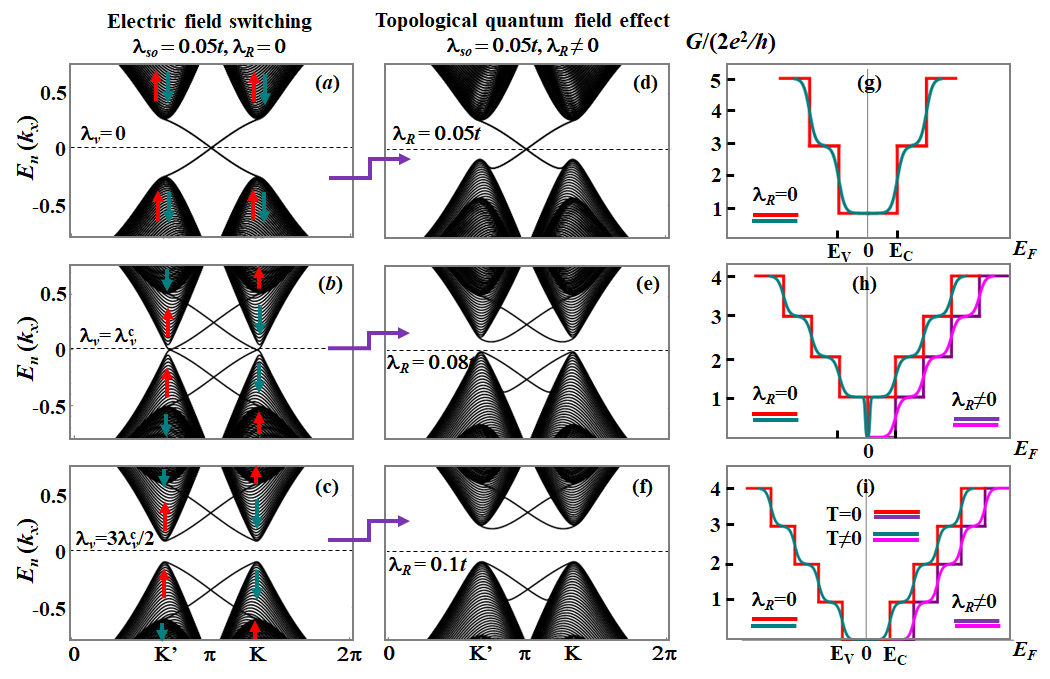}
    \caption{\textbf{Edge state dispersion and conductance quantization for semi-infinite honeycomb strip with zigzag edges}. \textbf{(a-c)} Electric field switching via topological phase phase transition showing helical edge states as available conducting channels in the ON state (a), critical gapless phase (b) and the trivial insulator or OFF state (c). \textbf{(d-f)} Topological quantum field effect in the "ON" state reduces the band gap by shifting the valence band along energy energy axis (d). On the other hand, at the critical point (e) and in the "OFF" state (f), topological quantum field effect enhances the trivial band gap where the maximum of valence band remains pinned but the minimum of conduction band lifted along energy axis. \textbf{(g-i)} Quantized conductance in terms of the number of modes $M(E)$ available at a given energy for TQFET in the QSH phase (g), at the critical point $\lambda_{v}=\lambda_{v}^c$ (h) and in the OFF state (i). While conduction band minimum remains pinned in ON state(g), (h) and (i) also show the shift of conduction band minimum in OFF state, along Fermi energy, induced by topological quantum field effect. In the ON (OFF) state, conductance jumps from $2e^2/h$ (0) to $6e^2/h$ ($2e^2/h$) at the tip of conduction/valence bands lying at the energy $E_c (E_v)$ while the electric field switching (b, e, h) is based upon manipulating the minimal conductance between $2e^2/h$ and $0$ and topological quantum field effect enhances this process. Here we set \textit{t} = 1 eV and the spin chirality of edge states is same as shown for the tips of connecting Dirac cones.}
    \label{ChainH}
\end{figure}

We first consider a simple case of electric field switching via a topological phase transition in a topological transistor by assuming that Rashba SOI is negligibly small $\Delta_{R}\approx0$ and the spin is a good quantum number. The low-energy single-particle band dispersion in the vicinity of the Dirac points reads
\begin{equation}
    E(\textbf{q},\lambda_{R}=0)= \pm \sqrt{v_{F}^2 |\textbf{q}|^2+\Bigg|\Delta_{so}+\eta_{s}\eta_{\tau}\Delta_{v}/2\Bigg|}
\end{equation}
where $\eta_{s}=+(-)$ stands for the spin up(down) sector and $\eta_{\tau}=+(-)$ represent valley K(K\textprime). A spin/valley dependent band gap $E_{G}(\lambda_{R}=0)=|2\Delta_{so}+\eta_{s}\eta_{\tau}\Delta_{v}|$ opens at the corners of the BZ. It shows that the electric field opens/tunes band gaps at both valleys symmetrically but, due to broken spin-valley degeneracy as manifested in figure \ref{TQFE-TBM}, the spin gaps are asymmetric and valley dependent. At a critical point, $\Delta_{v}=\Delta_{v}^c=2\Delta_{so}$, the system becomes semi-metallic, with both valleys K(K\textprime) perfectly spin-polarized hosting spin down (up) gapless phases. Away from the critical gapless phase, the system remains insulating with a finite band gap $E_{G}(\lambda_{R}=0)\ne0$. The bulk-boundary correspondence, as shown in figure \ref{ChainH}, confirms that the band gap is topologically nontrivial (QSH phase) for $0<\Delta_{v}<2\Delta_{so}$ while becoming trivial when $\Delta_{v}>2\Delta_{so}$.\par

\begin{flushleft}
\textbf{Topological quantum field effect on sub-threshold swing}\\
\end{flushleft}
In the transistor geometry shown in figure \ref{TQFET}, gate electric field $E_z$ applied perpendicular to the plane of 2D topological insulator material contacted by source and drain breaks mirror symmetry $M_z$ and induces spin-mixing Rashba SOI. When electric field induced Rashba SOI is also taken into account as a perturbative effect, the derived expression for the band gap
\begin{equation}
    E_{G}(\lambda_{R}\ne0)= \Bigg| 2\Delta_{so}-\frac{\Delta_{v}}{2}-\sqrt{\frac{\Delta_{v}^2}{4}+4\Delta_{R}^2}\Bigg|
\end{equation}
shows that the critical value of the staggered potential for the QSH-to-trivial insulating topological phase transition $\Delta_{v}^c=2(\Delta_{so}^2-\Delta_{R}^2)/\Delta_{so}$ is decreased by $2\Delta_{R}^2/\Delta_{so}$. Both the staggered sublattice potential and accompanied Rashba SOI are linear in the gate electric field $E_z$ and can be simulated as $\Delta_v=\alpha_vE_z$ and $\Delta_R=\alpha_RE_z$ respectively. Here $\alpha_v$ and $\alpha_R$ depend upon the lattice geometry and material specific parameters. The reduced sub-threshold swing of topological transistor then can be written in terms of $\alpha_v$ and $\alpha_R$ via band gap $E_{G}$ variation with gate electric field $E_z$ as $S^* = [(1/\alpha_v)dE_G/d(E_z)]^{-1}$. In the absence of Rashba SOI, the reduced sub-threshold swing $S^*(\lambda_R=0)$ is restricted to unity - Boltzmann's tyranny. However, in the process of electric field switching, Rashba SOI also influences the topologically trivial/nontrivial band gap and hence the electric field driven topological phase transition and we denote this effect as the topological quantum field effect. By incorporating the Rashba effect, Boltzmann's tyranny can be overcome in topological transistors via quantum field effect as
\begin{equation}
    S^*(\lambda_R\ne0)=\Bigg[\frac{1}{2}+\sqrt{\frac{1}{4}+\Bigg(\frac{2\alpha_R}{\alpha_v}\Bigg)^2}\Bigg]^{-1}
\end{equation}

It shows that a reduced sub-threshold swing of $S^*<0.75$ can easily be achieved when $\alpha_R>\alpha_v/3$. In order to understand the material realization, reduced sub-threshold swing can be quantified via estimation of $\alpha_R$ by finding Slater-Koster inter-orbital hopping parameters. In terms of band theory, mixing between $\sigma$ and $\pi$ bands due to intrinsic atomic SOI and Stark effect leads to a finite Rashba SOI. Based on the \textit{sp} microscopic tight binding model and $2^{nd}$ order perturbation theory \cite{HongkiMin06,Geissler13}, the explicit expression for Rashba SOI in buckled Xenes reads $\Delta_{R}= (ez\xi/3sin\theta V_{sp\sigma})E_{z}$. Here \textit{e} is the electon charge, \textit{z} is the Stark matrix element which is proportional to the size of atom at the site \textit{i}, $\xi$ is the atomic SOI, $\theta$ is the buckling angle, and $V_{sp\sigma}$ is the Slater-Koster parameter corresponding to the $\sigma$ bond formed by the \textit{s} and \textit{p} orbitals.\par 

In summary, the impact of topological quantum field effect on the trivial/nontrivial band gaps, critical value of electric field for ON/OFF switching, and reduced sub-threshold swing $S^*$ of the TQFET can be simulated in terms of atomic SOI, lattice parameters, and Slater-Koster inter-orbital hopping parameters as  
\begin{equation}
    E_{G}= \Bigg| 2\Delta_{so}-ed_{z}E_{z}\Bigg(\frac{1}{2}+\sqrt{\frac{1}{4}+\Bigg(\frac{2z\xi}{3d_zsin\theta V_{sp\sigma}}\Bigg)^2}\Bigg)\Bigg|
\end{equation}
\begin{equation}
    E_{z}^c=\frac{2\Delta_{so}}{ed_z}\Bigg[\frac{1}{2}+\sqrt{\frac{1}{4}+\Bigg(\frac{2z\xi}{3d_zsin\theta V_{sp\sigma}}\Bigg)^2}\Bigg]^{-1}
\end{equation}
\begin{equation}
    S^*=\Bigg[\frac{1}{2}+\sqrt{\frac{1}{4}+\Bigg(\frac{2z\xi}{3d_zsin\theta V_{sp\sigma}}\Bigg)^2}\Bigg]^{-1}
\end{equation}

From these expressions, it is apparent that a number of interesting features are captured by the topological quantum field effect. Firstly, Rashba SOI plays a central role in the topological phase transition driven by the gate electric field: Rashba SOI reduces the nontrivial band gap opened by the intrinsic SOI and enhances the trivial band gap opened by gate electric field. Electric field switching driven by a topological phase transition and the effect of associated topological quantum field effect is shown in Figure \ref{TQFE-TBM}. As shown in Figure \ref{TQFE-TBM}(d-f), we vary the Rashba SOI while keeping the intrinsic SOI fixed. In the non-trivial regime, the conduction band minima (CBM) at valley K(K\textprime) remain insensitive to the Rashba SOI strength. However, Rashba SOI reduces the nontrivial band gap by raising the valence band maxima (VBM) at valley K(K\textprime) along the energy axis. On the other hand, in the trivial regime, the VBM at valley K(K\textprime) remain pinned at valley K(K\textprime) but Rashba SOI increases the trivial band gap by shifting the CBM along the energy axis. The topological quantum field effect is best depicted at the transition point of the TQFET where the critical electric field leads to a gapless phase \ref{TQFE-TBM}(b) while the Rashba SOI $\lambda_{R}\ne0$ opens a trivial band gap via topological quantum field effect \ref{TQFE-TBM}(e). In short, the Rashba spin splitting of the valence (conduction) bands, especially along the edges of the BZ K-M-K\textprime, decreases (increases) the band gap in the nontrivial (trivial) regime and speeds up the electric field switching. \par

Secondly, in the absence of Rashba SOI, the nontrivial (trivial) band gap decreases (increases) linearly with the electric field and the reduced sub-threshold swing remains constant $S^*=1$. However, in the presence of Rashba SOI, up to leading order in atomic SOI $\xi$, the reduced sub-threshold swing decreases with increasing Rashba SOI strength and can be smaller than one, $S^*<1$. In addition, due to its dependence on the geometric structure of QSH lattices, atomic SOI, and the Slater-Koster inter-orbital hopping parameters, the topological quantum field effect provides tunable parameters for controlling the sub-threshold swing in a TQFET – rather than relying purely on the gate capacitance mechanism. \par

\begin{figure}[H]
    \centering
    \includegraphics[scale=0.75]{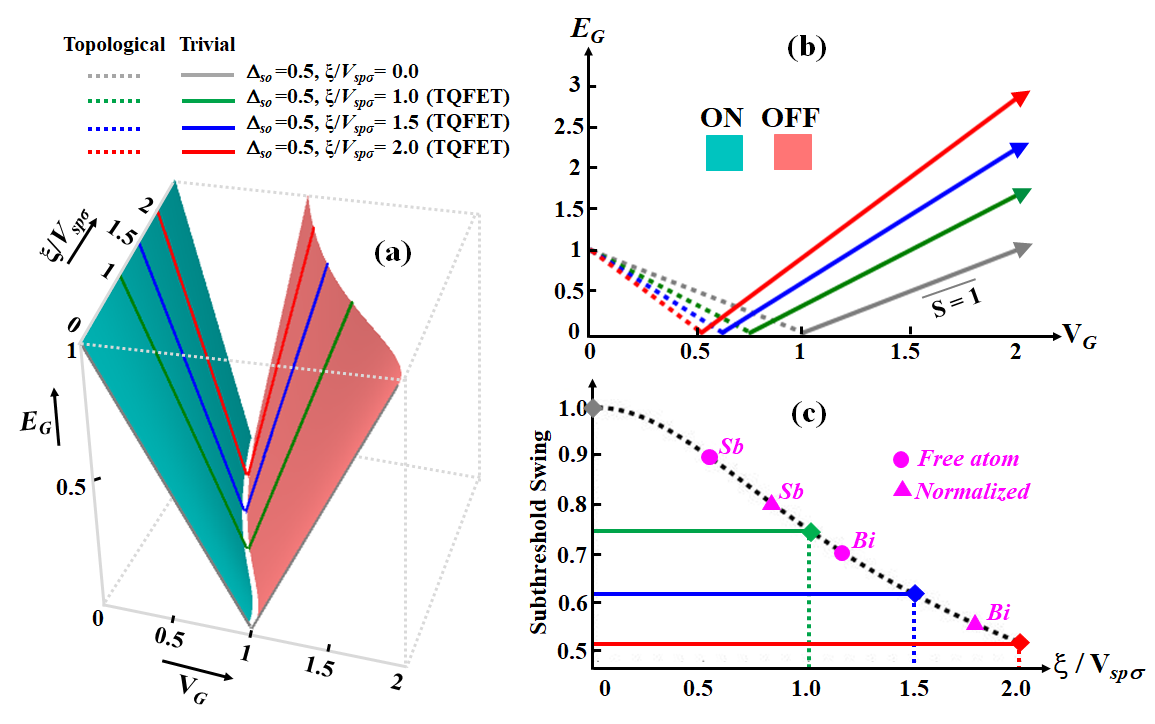}
    \caption{\textbf{Topological quantum field effect on band gap, threshold gate voltage, and sub-threshold swing.} The green, blue, and red lines represent the variation of band gap (\textbf{a}), threshold gate voltage (\textbf{b}), and sub-threshold swing (\textbf{c}) corresponding to atomic SOI and Slater-Koster parameter ratio $\xi/V_{sp\sigma}= 1, 1.5, 2$ respectively which encodes the topological quantum field effect. (\textbf{a}) Nontrivial (trivial) bulk band gap $E_G$ decreases (increases) sharply with increasing $\xi/V_{sp\sigma}$. Accordingly, the threshold gate voltage (\textbf{b}) and sub-threshold swing (\textbf{c}) decreases with increasing $\xi/V_{sp\sigma}$. Magenta circles (triangles) represent the sub-threshold swing for TQFET based on antimonene and bismuthene with free atomic (normalized) SOI. Here we assume that $d_{z}\approx z$ and $sin\theta\approx 1$ for quasi-planar/low-buckled honeycomb lattice.}
    \label{ST}
\end{figure}

The microscopic orbital picture shows that, for a TQFET based on a quasi-planar/low-buckled honeycomb lattice where $d_{z}\approx z$ and $sin\theta\approx 1$, both the threshold gate voltage and the sub-threshold swing decrease with increasing atomic SOI and Slater-Koster parameter ratio $\xi/V_{sp\sigma}$. As shown in Figure \ref{ST} and table \ref{tab:1}, 25\% reduction in  Boltzmann's limit of sub-threshold swing ($S^*<0.75$) can readily be achieved when $\xi/V_{sp\sigma}=1$. These constraints can be realized realistically in quasi-planar/low-buckled bismuthene sheets with band gap lying at K/K' valleys. Furthermore, in functionalized bismuth monolayers $BiX$ \cite{Song14} and $Bi_2XY$ \cite{Zhou18} where X/Y = H, F, Cl and Br, sub-threshold swing can be decreased by approximately 50\% ($S^*\approx0.5$) when $\xi/V_{sp\sigma}\approx2$ due to enhance atomic SOI as discussed below.\par

More importantly, the derived expression for suthreshold swing emphasizes that it has no fundamental lower bound and could be further improved by targeted materials design.For example, sub-threshold swing can be further reduced by enhancing another ratio $z/d_zsin\theta$ which purely depends on the nature of orbitals and geometric structure of 2D QSH lattices. One obvious path to further reduce sub-threshold swing is to enhance Stark matrix element $<\Phi_{n,l,m_l}|z_i|\Phi_{n,l\pm1,m_l}>$ at site \textit{i} by increasing transition between $s$ and $p_z$ orbitals via externally applied strain or substrate effect. Another possibility is to reduce factor $d_zsin\theta$ via optimization of geometric/buckling structure. Moreover, the  geometric structure of 2D QSH lattices can also be employed to tune other related parameters for electric field switching. For example, the first-order Kane-Mele type intrinsic SOI depends strongly on the buckling parameter $\theta$ and vanishes for planar honeycomb $\theta=90$. Such a geometric dependence of the intrinsic SOI also affects the topologically nontrivial band gap opened by second-order Kane-Mele type intrinsic SOI and hence the critical electric field for ON/OFF switching.\par

The reduction of sub-threshold swing stems from purely microscopic quantum phenomenon and is associated with bulk band topology. For example, FET based on Semenoff type \cite{Semenoff} topologically trivial insulating system where $\Delta_{so}=0$, the absolute trivial band gap opened by the gate electric field remains insensitive to the Rashba SOI which is to be contrasted with a TQFET based on QSH materials. Although topological quantum field effect in TQFET based on QSH insulators is endebted to $nnn$ intrinsic SOI $\Delta_{so}$ and hence the bulk band topology, the derived expression for the sub-threshold swing seems to remain independent of $\Delta_{so}$. However, the working of TQFET depends heavily on the nontrivial band gap opened by $\Delta_{so}$: a large intrinsic SOI $\Delta_{so}$ also requires a large gate electric field for topological switching, which automatically raises the strength of the associated Rashba SOI near the energy-zero TQFET operational regime. As shown by Min \textit{et. al.,} \cite{HongkiMin06} using a microscopic tight binding model, the strength of $\Delta_{R}$ ($\Delta_{so}$) for graphene is 100 times greater (smaller) than the value found by Kane and Mele based on symmetry analysis \cite{Kane05a,Kane05b}. This effect could be more prominent at large gate electric fields for heavy elemental group-V Xenes, especially antimonene and bismuthene with large atomic SOI as shown in Figure \ref{ST}. To realize a topologically nontrivial band gap and electric field switching between the QSH and trivial insulator, we require $\Delta_{so} > \Delta_{R}$. Otherwise, a honeycomb lattice becomes a zero-gap semiconductor when $\Delta_{so}=\Delta_{R}$ and a trivial metal for $\Delta_{so}<\Delta_{R}$.\par

Following a simple Landauer approach, the minimal quantized conductance can be obtained by finding the solution of the low energy effective tight-binding Hamiltonian for QSH honeycomb nanoribbons with zigzag edges. Close to equilibrium, $\mu_1=\mu_2=E_F$ where $\mu_{1(2)}$ is the chemical potential at the left (right) contact and $E_F$ is the Fermi energy which is controllable via doping, the conductance of a honeycomb nanoribbon with zigzag edges is calculated as a function of $E_F$. Figure \ref{ChainH} displays the conductance quantization at gate electric field $E_z =0$, $E_z=E_{z}^c$, and $E_z>E_{z}^c$ which correspond to "ON", "transition point", and "OFF" state of TQFET respectively.\par

In the QSH regime, at half filling, topologically protected edge states connecting opposite valleys cross the energy-zero at the time-reversal invariant momentum. With increasing $E_F$, the Fermi level crosses new bands and opens new channels for conductance. That is, as shown in figure \ref{ChainH}(b), the low-energy modes representing edge states (\textit{n} = 0) are valley non-degenerate while all the high-energy modes representing bulk states (\textit{n} = 1, 2, 3, . . .) are twofold valley-degenerate. As a result, the number of transverse modes available at energy \textit{E} can be expressed as $M(E) = 2n+1$. In addition, all the modes available for conductance retain a twofold spin degeneracy. Since each mode (for each spin and valley degree of freedom) acts as a channel that contributes to the conductance by $e^2/h$, the low-bias and low-temperature quantized conductance in the "ON" state of a TQFET can be expressed as $G_{ON}=(2e^2/h)(2n+1)\Tilde{T}$. Here $\Tilde{T}$ is the corresponding transmission probability per mode which goes to unity in ballistic QSH regime. As shown in figure \ref{ChainH}(c), the conductance plateau for $G_{ON}/(2e^2/h)$ appears at the "odd" integer values.\par

To estimate the conductance in the OFF state we adopt the Landauer approach again, assuming the channel to be connected to semiconducting leads so that the gap to transport is $E_G$. In the trivial insulating regime, as shown in Figure \ref{ChainH}(h), the near zero-energy minimum conducting channels vanishes due to dominating staggered potential term. Additionally, in the presence of SOI, the electric field also lifts the spin degeneracy such that the total number of conducting modes available at low energy are $M(E) = 2n$. As a result,the zero-temperature conductance in the trivial regime can be approximated as $G_{OFF}=(2e^2/h)n\Tilde{T}$. Note that, unlike spin-degenerate spectrum in QSH phase, the factor of 2 in conductance $G_{\textit{OFF}}$ is due to valley degrees of freedom and the corresponding conductance plateau for $G_{OFF}/(2e^2/h)$ appears at integer values \textit{n}=0,1,2,3… as shown in Figure \ref{ChainH}(i). Since we are interested here in manipulating the minimal conductance quantum, as the working of a TQFET deals with the shift from conductance quantum $2e^2/h$ to 0, the opted Landauer approach in trivial phase is also a reasonable approximation as long as the system is at half filling and the Fermi-level lies within the band gap $E_G$ of the nanoscale TQFET.\par

By using the Landauer formula, where the total electron transmission probability $M(E)\Tilde{T}$ at energy $E$ is convoluted with the energy derivative of the Fermi function $df(E)/dE$, the conductance $G_{OFF}$ of the OFF state is also plotted as a function of the Fermi energy at finite temperature. As shown in Figure \ref{ChainH}(i), the conductance plateaus becomes smoother for $T>0$. As expected, we also noted that the topological quantum field effect shifts the conductance plateau of $G_{OFF}$ along the energy axis as shown in Figure \ref{ChainH}(i): It is consistent with the band evolution for both infinite sheet as shown in \ref{TQFE-TBM} and semi-infinite sheet as shown in \ref{ChainH}(d): In the OFF state, the topological quantum field effect enhances the trivial band gap where the maximum of the valence band remains pinned but the minimum of the conduction band is lifted along the energy axis.

\begin{flushleft}
\textbf{Material realization and future directions}\\
\end{flushleft}
In general, both group-IV \cite{HongkiMin06,LiuPRL11,LiuPRB11,XuPRL13} and group-V \cite{Hsu15,Reis17,Li18} Xenes are QSH insulators and can serve the purpose of channel in topological FET. However, as shown in table \ref{tab:1}, weak Rashba SOI in group-IV Xenes has negligibly small effect on the sub-threshold swing. On the other hand, group-V Xenes with large atomic SOI such as bismuthene \cite{Hsu15,Reis17,Li18} or functionalized bismuth monolayers $BiX$ \cite{Song14} and $Bi_2XY$ \cite{Zhou18} where X/Y = H, F, Cl and Br, are promising materials for realizing topological quantum field effect in the proposed transistor geometry.\par

It has been shown that all the various group-V monolayer structures such as free-standing (As, Sb, Bi) monolayers, (As, Sb, Bi) monolayers on SiC substrate, and functionalized bismuth monolayers $BiX$ and $Bi_2XY$ are large gap QSH insulators. However, our framework developed for TQFET with reduced suthreshold swing ($S^*<0.75$) and high ON/OFF ratio ($10^5\sim10^{10}$) with large band gap in the OFF state ($>$10kT or 250meV) for functioning at room temperature highly desires buckled structure with bulk band gap away from the centre of Brillouin zone $\Gamma$. While buckling is required to realize electric field switching via topological quantum field effect, band gap at the corners of the Brillouin zone K(K\textprime) are highly desired for large bulk-band splitting via topological quantum field effect: Rashba SOI efficiently tunes the band gap by splitting the top/bottom of the valence/conduction bands lying at K(K\textprime) but has no effect on the band gap opened at the time-reversal invariant momenta. \par

While band gap in all the group-IV Xene structures, planar graphene and buckled silicene, germanene, and stanene, lies at the valleys K(K\textprime), location of Dirac points and the band gap in group-V Xenes depends upon the geometric structure, substrate effect, and their functionalization. For example, the Dirac points and the band gap lie at the high symmetry $\Gamma$-point in free-standing (As, Sb, Bi) monolayers with buckled honeycomb structures. However, (As, Sb, Bi)/SiC stabilizes in a planar honeycomb structure and hosts Dirac points and the band gap at the corners of the Brillouin zone \cite{Reis17,Li18}. On the other hand, functionalized bismuth monolayers $BiX$ and $Bi_2XY$ are ideal candidates where low-buckling is iduced by X/Y atoms while the Dirac points or the bulk band gap lies at the valleys K(K\textprime). Table \ref{tab:1} summarizes the strength of atomic SOI, Slater-Koster parameter $V_{sp\sigma}$ and sub-threshold swing in the presence of Rashba SOI for group-IV and V Xenes. \par

\begin{table}[H]
\setlength{\arrayrulewidth}{0.5mm}
\setlength{\tabcolsep}{5pt}
\renewcommand{\arraystretch}{1.5}
\begin{center}
\begin{tabular}{ |c|c|c|c|c|c| } 
\hline
\multirow{2}{2em}{}& \multicolumn{3}{c|}{$\xi(eV)$}&$V_{sp\sigma}$&$S^{\star\ddag}$\\
\cline{2-6}
&Free atom & Normalized & $Experimental^\dagger$ &(eV)&$\Delta_{R}\ne0$ \\
\hline
Graphene  &$0.006^a$&$0.009^k$& - &$5.580^f$ & 0.999[0.999] \\
Silicene  &$0.028^b$&$0.044^c$&(-)$0.044^{cd}$ &$2.54^g$  &0.999[0.999] \\
Germanene &$0.2^b$  &$0.29^c$ &(0.2)$0.29^{cd}$&$2.36^g$  &0.996[0.993]\\
Stanene   &$0.6^b$  &$0.8^c$ &(0.48)$0.77^{cd}$&$1.953^h$ &0.961[0.934]\\
\hline
Arsenene  &$0.29^b(0.36^e)$ & $0.421^c$ &- &$1.275^i$     &0.978[0.955]\\
Antimonene&$0.6^e(0.8^b)$  &$0.973^c$  &-&$1.170^i$      &0.904[0.802]\\
Bismuthene&$1.5^e$ & 2.25 &-&$1.3^{ij}$ &0.707[0.568]\\
\hline
\end{tabular}
\end{center}
\vspace{-0.5cm}
\caption{\label{tab:1} Strength of atomic SOI $\xi$, Slater-Koster parameter $V_{sp\sigma}$ and sub-threshold swing $S^*$ for TQFET based on group-IV and V Xenes. Here we assume that $d_{z}\approx z$ and $sin\theta\approx 1$ for quasi-planar/low-buckled honeycomb lattice. Similar to other group-IV and V elements, a normalization factor of 3/2 is also multiplied to bismuthene free atomic SOI.}
\end{table}
\vspace{-0.7cm}
\begin{flushleft}
a:Reference \cite{HongkiMin06} \hspace{0.4cm} 
b:Reference \cite{BRAUNSTEIN} \hspace{0.4cm}
c:Reference \cite{Chadi} \hspace{0.4cm}
d:Reference \cite{GIV-SOI-Exp} \hspace{0.4cm}
e:Reference \cite{Gonze-AsSbBi,Falicove-AsSbBi} \\
f:Reference \cite{Saito-G-TBM} \hspace{0.45cm}
g:Reference \cite{harrison1989electronic} \hspace{0.45cm}
h:Reference \cite{Liu-TBM-SbBiSn} \hspace{0.45cm}
i:Reference \cite{Xu-TBM-AsSbBi} \hspace{0.45cm}
j:Reference \cite{Chen-TBM-Bi}\\
k:Reference \cite{Yao-G-SOI}\\
$^\dagger$Entries in parentheses represent the experimental values for neutral atomic SOI. For experimental data, see reference \cite{Chadi,GIV-SOI-Exp} and references cited there in.\\
$^\ddag$Entries in brackets represent the sub-threshold swing for normalized SOI.
\end{flushleft}

Finally, the free atomic SOI $\xi=1.25$ for bimuthene sheets can be further enhanced in functionalized bismuth monolayers $BiX$ and $Bi_2XY$. Originally it was shown by Braunstein and Kane \cite{BRAUNSTEIN} that the spin-orbit splitting in free Ge atoms is 0.2eV while that in the solid it is 0.29eV \cite{GIV-SOI-Exp}. Similarly, Chadi \textit{et.al.,} showed that a renormalization factor of 3/2 works well for other group-IV and V structures \cite{Chadi}. Hence we considered both the free atomic $\xi_{0}=1.5eV$ and normalized SOI $\xi_{N}=3\xi_{0}/2=2.25eV$ for bismuthene as listed in \ref{tab:1}. Such a large renormalization can be understood as follows: The SOI term $(\hbar/4m^2c^2)(\mathbf{\nabla} V\times \mathbf{p}).\boldsymbol{\sigma}$ reduces to the well-known form $\xi(r)\mathbf{l}.\mathbf{s}$ where $\xi(r)=(\hbar^2/4m^2c^2)(\partial V/\partial r)$ when \textit{V} is approximated to be spherically symmetric. In terms of Wannier functions $w_{k}^s(r-d)$ centred about each lattice point whose vector coordinates are denoted by \textbf{d}, the SOI matrix elements are approximated as \cite{Elliott-SOI}\\
\begin{equation}
    \langle\psi_{s}(\mathbf{k})|\frac{\hbar}{4m^2c^2}(\mathbf{\nabla}V(\mathbf{r}-\mathbf{d})\times \mathbf{p}).\boldsymbol{\sigma}|\psi_{t}(\mathbf{k\textprime})\rangle = \delta_{kk\textprime}\langle w_{k}^s(r)|\xi(r)\mathbf{l}.\mathbf{s}|w_{k}^t(r)\rangle
\end{equation}
where wavefunctions $\psi_{s}(\mathbf{k})$ are superposition of Wannier functions and $V(\mathbf{r}-\mathbf{d})$ is the potential with a centre of symmetry at each lattice point. In tight binding methods, $\xi(r)$ is approximated as the strength of SOI under various constraints such that (i) The Wannier functions centred about different lattice points are orthogonal and hence localized in the Wigner-Seitz cell; (ii) $V(\mathbf{r}-\mathbf{d})$ centred about different lattice points possess spherical symmetry and is the same in each Wigner-Seitz cell and vanishes outside. However, the Wannier functions are not fully orthogonal and the potential \textit{V} does not have full spherical symmetry because of the surrounding atoms in a solid. Hence, the spilling over of Wannier functions into the nearest cells and the lower symmetry of \textit{V} lead to a substantial modification/renormalization.\par

The SOI renormalization is highly dependent on the various ionic configurations of group-IV and V elements in diamond and zinc-blende compounds \cite{BRAUNSTEIN,Chadi}. Due to the partial ionic character of chemical bonding, the spin-orbit splitting of cation \textit{p}-states in the compounds of group-IV and V elements depends upon the time spent by an electron around cation and anion sites. So the SOI of the cation \textit{p}-states in the group IV and V compounds, if displaying a similar QSH phase and a QSH to trivial insulator transition, can be further enhanced for quantum devices. The SOI dependence upon various ionic configurations for Sb ions is shown in Ref.~\cite{BRAUNSTEIN}. Following these lines, even for purely group-IV and V elemental honeycomb lattices, the atomic functions/orbitals near the top of valence bands need not be purely \textit{p}-like but can have admixtures of higher angular momentum (\textit{d}-like) states. That is, the renormalized SOI of cation p-states can be increased by either (i) promoting \textit{s} and \textit{p} electrons to higher states, or (ii) completely removing \textit{s} and \textit{p} electrons as in the ion. It can be achieved either through shining light or though gate controlled tuning of the energy of electronic states whose eigenvectors are non-vanishing at the singularities of the potential $V(\mathbf{r}-\mathbf{d})$.

\begin{flushleft}
\textbf{Conclusion}\\
\end{flushleft}
We have analysed the working of TQFET, employing the energy-zero edge state of QSH honeycomb nanoribbons, based on the conductance which is quantized, topologically protected, and indebted to the intrinsic microscopic quantum phenomena such as SOI and band topology. We noted that, associated with SOI and band topology and hence contrary to conventional semiconductor such as MOSFET or bilayer graphene, topological quantum field effect enhances the electric field switching and reduces the sub-threshold swing without any lower bound. This is subject to the topological quantum field effect modelled via Rashba SOI which provides tunable parameters for controlling the electric field switching and sub-threshold swing, in stark contrast to gate capacitance mechanism in conventional semiconductor transistor where sub-threshold swing is bounded by Boltzmann's tyranny. The derived expression for bulk band gap, critical electric field, and sub-threshold swing explicitly demonstrate that the working of TQFET can be controlled via geometric structure of QSH lattices, atomic SOI, and the Slater-Koster inter-orbital hopping parameters.\par

While incorporating the topological quantum field effect in engineering TQFET, a decrease of more than 25\% compared to Boltzmann's limit of sub-threshold swing is possible when $\xi/V_{sp\sigma}>1$ which can be realized in a bismuthene like staggered honeycomb structures. The sub-threshold swing can be further reduced through lattice distortion via strain and substrate effect: tuning of Slater-Koster parameters and Stark matrix elements and optimizing the Rashba SOI and buckling parameters. It shows that, unlike conventional semiconductor transistors, a TQFET does not have sharp lower bound on its sub-threshold swing and hence no \textit{topological tyranny}. In summary, topological quantum field effect, an alternate mechanism for reducing the sub-threshold swing, provides a promising platform for further research and developing energy-efficient quantum devices.\par

In passing, the topological quantum field effect can be employed  for simulating substrate effects in experimental condensed matter physics. Theoretically modelling of Rashba SOI induced by interaction between sample and substrate is more difficult as compared to one due to gate electric field. TQFET incorporating the topological quantum field effect can be employed as a device for estimating/modelling substrate induced Rashba SOI: first measure the sub-threshold swing for free-hanging 2D sheet and then perform the measurement again while incorporating the substrate effects. The strength of the substrate-induced Rashba SOI can be simulated via reduction in the sub-threshold swing of the TQFET.\\

\begin{flushleft}
\textbf{Acknowledgements.} This research is supported by the Australian Research Council (ARC) Centre of Excellence in Future Low-Energy Electronics Technologies (FLEET Project No. CE170100039) and funded by the Australian Government. 
M.N. and X.W. acknowledges the support from an ARC Professional Future Fellowship (FT130100778). 
M.S.F. acknowledges the support of an ARC Laureate Fellowship FL120100038. \\
\end{flushleft}

\printbibliography

@article{WaryL,
  title = {Topological transistor},
  author = {Wray, L. Andrew},
  journal = {Nature Physics},
  volume = {8},
  issue = {10},
  pages = {705-706},
  numpages = {2},
  year = {2012},
  month = {Oct},
  publisher = {Nature},
  doi = {10.1038/nphys2410},
  url = {https://doi.org/10.1038/nphys2410}
}

@article{Ezaw13APL,
author = {Ezawa,Motohiko },
title = {Quantized conductance and field-effect topological quantum transistor in silicene nanoribbons},
journal = {Applied Physics Letters},
volume = {102},
number = {17},
pages = {172103},
year = {2013},
doi = {10.1063/1.4803010},
URL = {https://doi.org/10.1063/1.4803010},
eprint = {https://doi.org/10.1063/1.4803010}
}

@article{Liu14-natmat,
  title = {Spin-filtered edge states with an electrically tunable gap in a two-dimensional topological crystalline insulator},
  author = {Liu, Junwei and Hsieh, Timothy H. and Wei, Peng and Duan, Wenhui and Moodera, Jagadeesh and Fu, Liang},
  journal = {Nature Materials},
  volume = {13},
  issue = {2},
  pages = {178-183},
  numpages = {5},
  year = {2014},
  month = {Feb},
  publisher = {Nature},
  doi = {10.1038/nmat3828},
  url = {https://doi.org/10.1038/nmat3828}
}

@article{Liu15-nano,
author = {Liu, Qihang and Zhang, Xiuwen and Abdalla, L. B. and Fazzio, Adalberto and Zunger, Alex},
title = {Switching a Normal Insulator into a Topological Insulator via Electric Field with Application to Phosphorene},
journal = {Nano Letters},
volume = {15},
number = {2},
pages = {1222-1228},
year = {2015},
doi = {10.1021/nl5043769},
    note ={PMID: 25607525},
URL = {https://doi.org/10.1021/nl5043769},
eprint = {https://doi.org/10.1021/nl5043769}
}

@article{pan15sci,
  title = {Electric control of topological phase transitions in Dirac semimetal thin films},
  author = {Pan, Hui and Wu, Meimei and Liu, Ying and Yang, Shengyuan A.},
  journal = {Scientific Reports},
  volume = {5},
  issue = {1},
  pages = {14639},
  numpages = {7},
  year = {2015},
  month = {Sep},
  publisher = {Nature},
  doi = {10.1038/srep14639},
  url = {https://doi.org/10.1038/srep14639}
}

@article {Qian14,
	author = {Qian, Xiaofeng and Liu, Junwei and Fu, Liang and Li, Ju},
	title = {Quantum spin Hall effect in two-dimensional transition metal dichalcogenides},
	volume = {346},
	number = {6215},
	pages = {1344--1347},
	year = {2014},
	doi = {10.1126/science.1256815},
	publisher = {American Association for the Advancement of Science},
	issn = {0036-8075},
	URL = {https://science.sciencemag.org/content/346/6215/1344},
	eprint = {https://science.sciencemag.org/content/346/6215/1344.full.pdf},
	journal = {Science}
}

@article{zhange-natnano17,
  title = {Magnetic quantum phase transition in Cr-doped $Bi_{2}(Se_{x} Te_{1-x})_{3}$ driven by the Stark effect},
  author = {Zhang, Zuocheng and Feng, Xiao and Wang, Jing and Lian, Biao and Zhang, Jinsong and Chang, Cuizu and Guo, Minghua and Ou, Yunbo and Feng, Yang and Zhang, Shou-Cheng and He, Ke and Ma, Xucun and Xue, Qi-Kun and Wang, Yayu},
  journal = {Nature Nanotechnology},
  volume = {10},
  issue = {12},
  pages = {953-957},
  numpages = {5},
  year = {2017},
  month = {Oct},
  publisher = {Nature},
  doi = {10.1038/nnano.2017.149},
  url = {https://doi.org/10.1038/nnano.2017.149}
}

@article{Molle-natmat17,
  title = {Buckled two-dimensional Xene sheets},
  author = {Molle, Alessandro and Goldberger, Joshua and Houssa, Michel and Xu, Yong and Zhang, Shou-Cheng and Akinwande, Deji},
  journal = {Nature Materials},
  volume = {16},
  issue = {02},
  pages = {163-169},
  numpages = {57},
  year = {2017},
  month = {Feb},
  publisher = {Nature},
  doi = {10.1038/nmat4802},
  url = {https://doi.org/10.1038/nmat4802}
}

@article{Collins18,
  title = {Electric-field-tuned topological phase transition in ultrathin $Na_{3}Bi$},
  author = {Collins, James L. and Tadich, Anton and Wu, Weikang and Gomes, Lidia C. and Rodrigues, Joao N. B. and Liu, Chang and Hellerstedt, Jack and Ryu, Hyejin and Tang, Shujie and Mo, Sung-Kwan and Adam, Shaffique and Yang, Shengyuan A. and Fuhrer, Michael S. and Edmonds, Mark T.},
  journal = {Nature},
  volume = {564},
  issue = {7736},
  pages = {390-394},
  numpages = {5},
  year = {2018},
  month = {Dec},
  publisher = {Nature},
  doi = {10.1038/s41586-018-0788-5},
  url = {https://doi.org/10.1038/s41586-018-0788-5}
}

@article{Kane05a,
  title = {${Z}_{2}$ Topological Order and the Quantum Spin Hall Effect},
  author = {Kane, C. L. and Mele, E. J.},
  journal = {Phys. Rev. Lett.},
  volume = {95},
  issue = {14},
  pages = {146802},
  numpages = {4},
  year = {2005},
  month = {Sep},
  publisher = {American Physical Society},
  doi = {10.1103/PhysRevLett.95.146802},
  url = {https://link.aps.org/doi/10.1103/PhysRevLett.95.146802}
}

@article{Kane05b,
  title = {Quantum Spin Hall Effect in Graphene},
  author = {Kane, C. L. and Mele, E. J.},
  journal = {Phys. Rev. Lett.},
  volume = {95},
  issue = {22},
  pages = {226801},
  numpages = {4},
  year = {2005},
  month = {Nov},
  publisher = {American Physical Society},
  doi = {10.1103/PhysRevLett.95.226801},
  url = {https://link.aps.org/doi/10.1103/PhysRevLett.95.226801}
}

@article{Rashba09,
  title = {Graphene with structure-induced spin-orbit coupling: Spin-polarized states, spin zero modes, and quantum Hall effect},
  author = {Rashba, Emmanuel I.},
  journal = {Phys. Rev. B},
  volume = {79},
  issue = {16},
  pages = {161409},
  numpages = {4},
  year = {2009},
  month = {Apr},
  publisher = {American Physical Society},
  doi = {10.1103/PhysRevB.79.161409},
  url = {https://link.aps.org/doi/10.1103/PhysRevB.79.161409}
}

@article{HongkiMin06,
  title = {Intrinsic and Rashba spin-orbit interactions in graphene sheets},
  author = {Min, Hongki and Hill, J. E. and Sinitsyn, N. A. and Sahu, B. R. and Kleinman, Leonard and MacDonald, A. H.},
  journal = {Phys. Rev. B},
  volume = {74},
  issue = {16},
  pages = {165310},
  numpages = {5},
  year = {2006},
  month = {Oct},
  publisher = {American Physical Society},
  doi = {10.1103/PhysRevB.74.165310},
  url = {https://link.aps.org/doi/10.1103/PhysRevB.74.165310}
}

@article{Geissler13,
	doi = {10.1088/1367-2630/15/8/085030},
	url = {https://doi.org/10.1088%2F1367-2630%2F15%2F8%2F085030},
	year = 2013,
	month = {aug},
	publisher = {{IOP} Publishing},
	volume = {15},
	number = {8},
	pages = {085030},
	author = {F Geissler and J C Budich and B Trauzettel},
	title = {Group theoretical and topological analysis of the quantum spin Hall effect in silicene},
	journal = {New Journal of Physics}
}

@article{LiuPRL11,
  title = {Quantum Spin Hall Effect in Silicene and Two-Dimensional Germanium},
  author = {Liu, Cheng-Cheng and Feng, Wanxiang and Yao, Yugui},
  journal = {Phys. Rev. Lett.},
  volume = {107},
  issue = {7},
  pages = {076802},
  numpages = {4},
  year = {2011},
  month = {Aug},
  publisher = {American Physical Society},
  doi = {10.1103/PhysRevLett.107.076802},
  url = {https://link.aps.org/doi/10.1103/PhysRevLett.107.076802}
}

@article{LiuPRB11,
  title = {Low-energy effective Hamiltonian involving spin-orbit coupling in silicene and two-dimensional germanium and tin},
  author = {Liu, Cheng-Cheng and Jiang, Hua and Yao, Yugui},
  journal = {Phys. Rev. B},
  volume = {84},
  issue = {19},
  pages = {195430},
  numpages = {11},
  year = {2011},
  month = {Nov},
  publisher = {American Physical Society},
  doi = {10.1103/PhysRevB.84.195430},
  url = {https://link.aps.org/doi/10.1103/PhysRevB.84.195430}
}

@article{XuPRL13,
  title = {Large-Gap Quantum Spin Hall Insulators in Tin Films},
  author = {Xu, Yong and Yan, Binghai and Zhang, Hai-Jun and Wang, Jing and Xu, Gang and Tang, Peizhe and Duan, Wenhui and Zhang, Shou-Cheng},
  journal = {Phys. Rev. Lett.},
  volume = {111},
  issue = {13},
  pages = {136804},
  numpages = {5},
  year = {2013},
  month = {Sep},
  publisher = {American Physical Society},
  doi = {10.1103/PhysRevLett.111.136804},
  url = {https://link.aps.org/doi/10.1103/PhysRevLett.111.136804}
}

@article{Hsu15,
	doi = {10.1088/1367-2630/17/2/025005},
	url = {https://doi.org/10.1088%2F1367-2630%2F17%2F2%2F025005},
	year = 2015,
	month = {feb},
	publisher = {{IOP} Publishing},
	volume = {17},
	number = {2},
	pages = {025005},
	author = {Chia-Hsiu Hsu and Zhi-Quan Huang and Feng-Chuan Chuang and Chien-Cheng Kuo and Yu-Tzu Liu and Hsin Lin and Arun Bansil},
	title = {The nontrivial electronic structure of Bi/Sb honeycombs on {SiC}(0001)},
	journal = {New Journal of Physics},
}

@article {Reis17,
	author = {Reis, F. and Li, G. and Dudy, L. and Bauernfeind, M. and Glass, S. and Hanke, W. and Thomale, R. and Sch{\"a}fer, J. and Claessen, R.},
	title = {Bismuthene on a SiC substrate: A candidate for a high-temperature quantum spin Hall material},
	volume = {357},
	number = {6348},
	pages = {287--290},
	year = {2017},
	doi = {10.1126/science.aai8142},
	publisher = {American Association for the Advancement of Science},
	issn = {0036-8075},
	URL = {https://science.sciencemag.org/content/357/6348/287},
	eprint = {https://science.sciencemag.org/content/357/6348/287.full.pdf},
	journal = {Science}
}

@article{Li18,
  title = {Theoretical paradigm for the quantum spin Hall effect at high temperatures},
  author = {Li, Gang and Hanke, Werner and Hankiewicz, Ewelina M. and Reis, Felix and Sch\"afer, J\"org and Claessen, Ralph and Wu, Congjun and Thomale, Ronny},
  journal = {Phys. Rev. B},
  volume = {98},
  issue = {16},
  pages = {165146},
  numpages = {15},
  year = {2018},
  month = {Oct},
  publisher = {American Physical Society},
  doi = {10.1103/PhysRevB.98.165146},
  url = {https://link.aps.org/doi/10.1103/PhysRevB.98.165146}
}

@article{Song14,
  title = {Quantum spin Hall insulators and quantum valley Hall insulators of BiX/SbX (X=H, F, Cl and Br) monolayers with a record bulk band gap},
  author = {Song, Zhigang and Liu, Cheng-Cheng and Yang, Jinbo and Han, Jingzhi and Ye, Meng and Fu, Botao and Yang, Yingchang and Niu, Qian and Lu, Jingand and Yao, Yugui},
  journal = {NPG Asia Materials},
  volume = {06},
  issue = {12},
  pages = {e147-e147},
  numpages = {7},
  year = {2014},
  month = {Dec},
  publisher = {Nature Publishing Group},
  doi = {10.1038/am.2014.113},
  url = {https://doi.org/10.1038/am.2014.113}
}

@article{Zhou18,
  title = {Giant spin-valley polarization and multiple Hall effect in functionalized bismuth monolayers},
  author = {Zhou, Tong and Zhang, Jiayong and Jiang, Hua and Žutić, Igor and Yang, Zhongqin},
  journal = {npj Quantum Materials},
  volume = {03},
  issue = {1},
  pages = {-},
  numpages = {7},
  year = {2018},
  month = {Aug},
  publisher = {Nature Publishing Group},
  doi = {10.1038/s41535-018-0113-4},
  url = {https://doi.org/10.1038/s41535-018-0113-4}
}

@article{BRAUNSTEIN,
title = "The valence band structure of the III–V compounds",
journal = "Journal of Physics and Chemistry of Solids",
volume = "23",
number = "10",
pages = "1423 - 1431",
year = "1962",
issn = "0022-3697",
doi = "https://doi.org/10.1016/0022-3697(62)90195-6",
url = "http://www.sciencedirect.com/science/article/pii/0022369762901956",
author = "R. Braunstein and E.O. Kane",
}

@article{Gonze-AsSbBi,
  title = {First-principles study of As, Sb, and Bi electronic properties},
  author = {Gonze, X. and Michenaud, J.-P. and Vigneron, J.-P.},
  journal = {Phys. Rev. B},
  volume = {41},
  issue = {17},
  pages = {11827--11836},
  numpages = {0},
  year = {1990},
  month = {Jun},
  publisher = {American Physical Society},
  doi = {10.1103/PhysRevB.41.11827},
  url = {https://link.aps.org/doi/10.1103/PhysRevB.41.11827}
}

@article{Falicove-AsSbBi,
  title = {Electronic Band Structure of Arsenic. I. Pseudopotential Approach},
  author = {Falicov, L. M. and Golin, Stuart},
  journal = {Phys. Rev.},
  volume = {137},
  issue = {3A},
  pages = {A871--A882},
  numpages = {0},
  year = {1965},
  month = {Feb},
  publisher = {American Physical Society},
  doi = {10.1103/PhysRev.137.A871},
  url = {https://link.aps.org/doi/10.1103/PhysRev.137.A871}
}

@article{Chadi,
  title = {Spin-orbit splitting in crystalline and compositionally disordered semiconductors},
  author = {Chadi, D. J.},
  journal = {Phys. Rev. B},
  volume = {16},
  issue = {2},
  pages = {790--796},
  numpages = {0},
  year = {1977},
  month = {Jul},
  publisher = {American Physical Society},
  doi = {10.1103/PhysRevB.16.790},
  url = {https://link.aps.org/doi/10.1103/PhysRevB.16.790}
}

@article{GIV-SOI-Exp,
  title = {Electroreflectance at a Semiconductor-Electrolyte Interface},
  author = {Cardona, Manuel and Shaklee, Kerry L. and Pollak, Fred H.},
  journal = {Phys. Rev.},
  volume = {154},
  issue = {3},
  pages = {696--720},
  numpages = {0},
  year = {1967},
  month = {Feb},
  publisher = {American Physical Society},
  doi = {10.1103/PhysRev.154.696},
  url = {https://link.aps.org/doi/10.1103/PhysRev.154.696}
}

@article{Saito-G-TBM,
  title = {Electronic structure of graphene tubules based on $C_{60}$},
  author = {Saito, Riichiro and Fujita, Mitsutaka and Dresselhaus, G. and Dresselhaus, M. S.},
  journal = {Phys. Rev. B},
  volume = {46},
  issue = {3},
  pages = {1804-1811},
  numpages = {8},
  year = {1992},
  month = {Jul},
  publisher = {American Physical Society},
  doi = {10.1103/PhysRevB.46.1804},
  url = {https://link.aps.org/doi/10.1103/PhysRevB.46.1804}
}

@book{harrison1989electronic,
  title={Electronic structure and the properties of solids: the physics of the chemical bond},
  author={Harrison, W.A.},
  isbn={9780486660219},
  lccn={89034153},
  series={Dover Books on Physics},
  url={https://books.google.com.au/books?id=orAPAQAAMAAJ},
  year={1989},
  publisher={Dover Publications}
}

@article{Liu-TBM-SbBiSn,
  title = {Electronic structure of the semimetals Bi and Sb},
  author = {Liu, Yi and Allen, Roland E.},
  journal = {Phys. Rev. B},
  volume = {52},
  issue = {3},
  pages = {1566--1577},
  numpages = {0},
  year = {1995},
  month = {Jul},
  publisher = {American Physical Society},
  doi = {10.1103/PhysRevB.52.1566},
  url = {https://link.aps.org/doi/10.1103/PhysRevB.52.1566}
}

@article{Xu-TBM-AsSbBi,
  title = {Tight-binding theory of the electronic structures for rhombohedral semimetals},
  author = {Xu, J. H. and Wang, E. G. and Ting, C. S. and Su, W. P.},
  journal = {Phys. Rev. B},
  volume = {48},
  issue = {23},
  pages = {17271--17279},
  numpages = {0},
  year = {1993},
  month = {Dec},
  publisher = {American Physical Society},
  doi = {10.1103/PhysRevB.48.17271},
  url = {https://link.aps.org/doi/10.1103/PhysRevB.48.17271}
}

@article{Chen-TBM-Bi,
	doi = {10.1088/1367-2630/aaca24},
	url = {https://doi.org/10.1088%2F1367-2630%2Faaca24},
	year = 2018,
	month = {jun},
	publisher = {{IOP} Publishing},
	volume = {20},
	number = {6},
	pages = {062001},
	author = {Szu-Chao Chen and Jhao-Ying Wu and Ming-Fa Lin},
	title = {Feature-rich magneto-electronic properties of bismuthene},
	journal = {New Journal of Physics},
}

@article{Yao-G-SOI,
  title = {Spin-orbit gap of graphene: First-principles calculations},
  author = {Yao, Yugui and Ye, Fei and Qi, Xiao-Liang and Zhang, Shou-Cheng and Fang, Zhong},
  journal = {Phys. Rev. B},
  volume = {75},
  issue = {4},
  pages = {041401},
  numpages = {4},
  year = {2007},
  month = {Jan},
  publisher = {American Physical Society},
  doi = {10.1103/PhysRevB.75.041401},
  url = {https://link.aps.org/doi/10.1103/PhysRevB.75.041401}
}

@article{Elliott-SOI,
  title = {Theory of the Effect of Spin-Orbit Coupling on Magnetic Resonance in Some Semiconductors},
  author = {Elliott, R. J.},
  journal = {Phys. Rev.},
  volume = {96},
  issue = {2},
  pages = {266--279},
  numpages = {0},
  year = {1954},
  month = {Oct},
  publisher = {American Physical Society},
  doi = {10.1103/PhysRev.96.266},
  url = {https://link.aps.org/doi/10.1103/PhysRev.96.266}
}

@article{Slater-Koster,
  title = {Simplified LCAO Method for the Periodic Potential Problem},
  author = {Slater, J. C. and Koster, G. F.},
  journal = {Phys. Rev.},
  volume = {94},
  issue = {6},
  pages = {1498--1524},
  numpages = {0},
  year = {1954},
  month = {Jun},
  publisher = {American Physical Society},
  doi = {10.1103/PhysRev.94.1498},
  url = {https://link.aps.org/doi/10.1103/PhysRev.94.1498}
}

@article{Shan10,
	doi = {10.1088/1367-2630/12/4/043048},
	url = {https://doi.org/10.1088%2F1367-2630%2F12%2F4%2F043048},
	year = 2010,
	month = {apr},
	publisher = {{IOP} Publishing},
	volume = {12},
	number = {4},
	pages = {043048},
	author = {Wen-Yu Shan and Hai-Zhou Lu and Shun-Qing Shen},
	title = {Effective continuous model for surface states and thin films of three-dimensional topological insulators},
	journal = {New Journal of Physics},
}

@article{SCZhnag,
  title = {Oscillatory crossover from two-dimensional to three-dimensional topological insulators},
  author = {Liu, Chao-Xing and Zhang, HaiJun and Yan, Binghai and Qi, Xiao-Liang and Frauenheim, Thomas and Dai, Xi and Fang, Zhong and Zhang, Shou-Cheng},
  journal = {Phys. Rev. B},
  volume = {81},
  issue = {4},
  pages = {041307},
  numpages = {4},
  year = {2010},
  month = {Jan},
  publisher = {American Physical Society},
  doi = {10.1103/PhysRevB.81.041307},
  url = {https://link.aps.org/doi/10.1103/PhysRevB.81.041307}
}

@article{Semenoff,
  title = {Condensed-Matter Simulation of a Three-Dimensional Anomaly},
  author = {Semenoff, Gordon W.},
  journal = {Phys. Rev. Lett.},
  volume = {53},
  issue = {26},
  pages = {2449--2452},
  numpages = {0},
  year = {1984},
  month = {Dec},
  publisher = {American Physical Society},
  doi = {10.1103/PhysRevLett.53.2449},
  url = {https://link.aps.org/doi/10.1103/PhysRevLett.53.2449}
}

@article{Kobayashi18,
	doi = {10.7567/apex.11.110101},
	url = {https://doi.org/10.7567%2Fapex.11.110101},
	year = 2018,
	month = {oct},
	publisher = {{IOP} Publishing},
	volume = {11},
	number = {11},
	pages = {110101},
	author = {Masaharu Kobayashi},
	title = {A perspective on steep-subthreshold-slope negative-capacitance field-effect transistor},
	journal = {Applied Physics Express}
}

@ARTICLE{Newns00,
  author={
  Newns, D. M. and Doderer, T. and Tsuei, C. C. and
Donath, W. M. and Misewich, J. A. and Gupta, A. and Grossman, B. M. and Schrott, A. and Scott, B. A. and Pattnaik, P. C. and
von Gutfeld, R. J. and Sun, J. Z.},
  journal={Journal of Electroceramics}, 
  title={The Mott Transition Field Effect Transistor: A Nanodevice?}, 
  year={2000},
  volume={4},
  pages={339–344}
  }

@ARTICLE{Banerjee09,
  author={S. K. {Banerjee} and L. F. {Register} and E. {Tutuc} and D. {Reddy} and A. H. {MacDonald}},
  journal={IEEE Electron Device Letters}, 
  title={Bilayer PseudoSpin Field-Effect Transistor (BiSFET): A Proposed New Logic Device}, 
  year={2009},
  volume={30},
  number={2},
  pages={158-160}
  }

@ARTICLE{Iniguez19,
  author={Íñiguez, Jorge and Zubko, Pavlo and Luk’yanchuk, Igor and Cano, Andrés},
  journal={Nature Reviews Materials}, 
  title={Ferroelectric negative capacitance}, 
  year={2019},
  volume={4},
  pages={243-256}
  }

@book{collaert2018,
  title={High Mobility Materials for CMOS Applications},
  author={Collaert, N.},
  isbn={9780081020623},
  series={Woodhead Publishing Series in Electronic and Optical Materials},
  url={https://books.google.com.au/books?id=sOJgDwAAQBAJ},
  year={2018},
  publisher={Elsevier Science}
}

@ARTICLE{Ionescu11,
  author={Ionescu, Adrian M. and Riel, Heike},
  journal={Nature}, 
  title={Tunnel field-effect transistors as energy-efficient electronic switches}, 
  year={2011},
  volume={479},
  pages={329-337}
  }

@article{McCann13,
	doi = {10.1088/0034-4885/76/5/056503},
	url = {https://doi.org/10.1088%2F0034-4885%2F76%2F5%2F056503},
	year = 2013,
	month = {apr},
	publisher = {{IOP} Publishing},
	volume = {76},
	number = {5},
	pages = {056503},
	author = {Edward McCann and Mikito Koshino},
	title = {The electronic properties of bilayer graphene},
	journal = {Reports on Progress in Physics},
}
\end{document}